\documentclass[final,5p,times,twocolumn]{elsarticle}

\usepackage{amssymb}
\usepackage{graphicx}

\begin{document}

\begin{frontmatter}


\title{Optimal Tree for Both Synchronizability and Converging Time}
\author[]{An Zeng}
\author[]{Yanqing Hu}
\author[]{Zengru Di\corref{cor1}}
\ead{zdi@bnu.edu.cn} \cortext[cor1]{Corresponding author}
\address{Department of Systems Science, School of Management, Beijing Normal University, Beijing 100875, P.R. China}

\begin{abstract}
It has been proved that the spanning tree from a given network has
the optimal synchronizability, which means the index
$R=\lambda_{N}/\lambda_{2}$ reaches the minimum 1. Although the
optimal synchronizability is corresponding to the minimal critical
overall coupling strength to reach synchronization, it does not
guarantee a shorter converging time from disorder initial
configuration to synchronized state. In this letter, we find that it
is the depth of the tree that affects the converging time. In
addition, we present a simple and universal way to get such an
effective oriented tree in a given network to reduce the converging
time significantly by minimizing the depth of the tree. The shortest
spanning tree has both the maximal synchronizability and efficiency.

\end{abstract}

\end{frontmatter}

\section{Introduction}
The synchronization is an universal phenomenon emerged by a
population of dynamically interacting units. It plays an important
role from physics to biology and has attracted much attention for
hundreds of years. Thus, there are a great deal of relative
researches based on this topic. With the understanding of relations
between network topology and the synchronizability[1-5], scientists
have proposed many methods to enhance synchronizability of the
network[6-15]. Some of them tried to modify the topology of the
network to enhance the synchronization[6-10] while others by just
modifying the coupling weight of each edge while keeping the
topology unchanged[11-16].

In these papers, the synchronization is always measured by the
eigenvalues of Laplacian matrix as
$R=\lambda_{N}/\lambda_{2}$[17-19]. The smaller the $R$ is, the
better the synchronizability will be. In recent years, a lot of
works focus on how to enhance the synchronizability by distributing
the weight to the edges according to the structural characteristics
of the nodes and edges. For example, distributing the weight by the
degree and betweeness can sharply enhance the network
synchronizability, these methods can reduce index $R$ to a small
value[11-14]. For the growing scale-free network, some researchers
took the age into consideration and reduced the index $R$ to an even
smaller value close to the minimum 1[15-16]. In the Ref[20],
Nishikawa and Motter gave the weight distributing an extreme way by
imposing the weight of some edges to 0. The process can be regarded
as cutting off the edges which are disadvantage to the
synchronization. Finally, they can get an oriented tree with
normalized input strength and no directed loops. Moreover, they
proved that the $R$ of the tree is 1, meaning that the tree has
maximal synchronizability. The index $R$ is corresponding to the
critical overall coupling strength. When $R$ reaches its minimum 1,
the synchronized states are stable for the widest possible range of
the parameter representing the overall coupling strength.

When investigating the synchronization of a network, we always
consider both of the critical overall coupling strength to
synchronized the whole network and the converging time. The
converging time can be regarded as the efficiency of a network.
However, better synchronizability does not guarantee a shorter
converging time from disorder initial configuration to synchronized
state. Actually, in the Ref[20], Nishikawa and Motter found that the
synchronizing process may take longer time in the optimal network
with $R=1$. It leads us to an interesting problem that what is the
factor that affects the converging time. In this paper, we will give
a clear answer to this problem. In addition, a simple and universal
method is presented to find the optimal spanning tree with maximal
synchronizability and efficiency. This kind of tree is the optimal
structure for synchronization that the scientists try to find from
any given network.

\section{Result}
\subsection{The factor affecting the converging time}
In a dynamical network, each node represents an oscillator and the
edges represent the couplings between the nodes. For a network of
$N$ linearly coupled identical oscillators, the dynamical equation
of each oscillator can be written as

\begin{equation}
\dot{x_{i}}=F(x_{i})-\sigma\sum\limits_{j=1}\limits^{N}G_{ij}H(x_{j}),\quad
\quad i=1,2.....,N
\end{equation}

It has been proved that the synchronizability of an oriented tree of
the network has reaches its maximum, that is the index $R=1$ and the
synchronized states are stable for the widest possible range of
overall coupling strength. Generally, this kind of oriented tree has
a root with no input. It works as the master oscillator and affects
the oscillators in the hierarchical level below without any
feedback. Then, the next lower level oscillators will get
synchronized and so on until the whole network reaches complete
synchronization. Hence, the hierarchy number is very important for
it determines the converging time[21]. Clearly, in this kind of
spanning tree, the synchronized process in any branch is independent
with each other. So the oscillator in the lowest hierarchical level
would reach the synchronized state at last. From this point of view,
we figure that when given the oscillator model and the overall
coupling strength, the converging time is only determined by the
depth of the tree.

In order to validate the assumption, we put the Kuramoto model to
each node of the network to  make numerical simulation. The Kuramoto
model is a classical model to investigate the phase synchronization
phenomenon[22-24]. The coupled Kuramoto model in the network can be
written as

\begin{equation}
\dot{\theta_{i}}=\omega_{i}+\sigma
G_{ij}sin(\theta_{j}-\theta_{i}),\quad \quad i=1,2.....,N
\end{equation}

The collective dynamics of the whole population is measured by the
macroscopic complex order parameter,

\begin{equation}
r(t)e^{i\phi(t)}=\frac{1}{N}\sum\limits_{j=1}\limits^{N}e^{i\theta_{j}(t)}
\end{equation}

Where the $r(t)\simeq1$ and $r(t)\simeq0$ describe the limits in
which all oscillators are either phase locked or move incoherently,
respectively. In this paper, we define the time $T$ as the
converging time, where for any $t>T$, $r(t)-r(t-dt)<\varepsilon$ all
along and the $\varepsilon=10^{-6}$.

\begin{figure}
  \center
  \includegraphics[width=8cm]{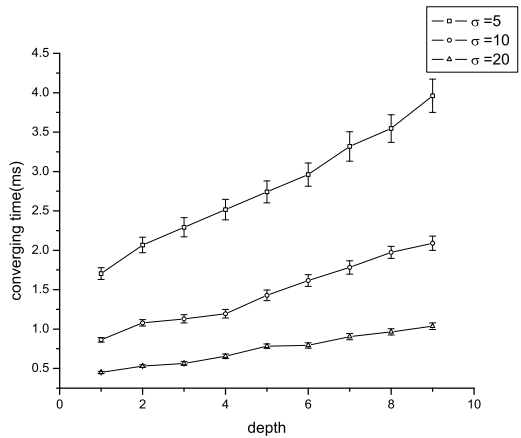}\\
  \caption{The converging time to synchronization under trees with different depth. All the trees are of the same size $N=10$.
   The only different between them is the depth of the tree. The $\square$ stands for the overall coupling strength $\sigma$ is 5, the
   $\bigcirc$ stands for $\sigma=10$ and the $\bigtriangleup$ stands for $\sigma=20$ .The result is averaged by
50 times.}
\end{figure}

To investigate which affects the converging time of the network, we
adopt the trees with 10 nodes. When the depth is 1, the tree has
only one root, all the other nodes which only receive input from the
root are located in the second hierarchical level. When the depth of
the tree is 9, the tree is just a chain connecting all the nodes.
When the depth of the tree is from 2 to 8, the structures are more
complicated.  We compare the converging time of the trees with
different depths. Since the trees with the same depth may have
different structures, we use the average $T$ to represent the
converging time. From Fig.1, it is clear that the deeper the tree
is, the more converging time is needed to reach synchronized state.
The effect of the depth to converging time is linear roughly.

Furthermore, given the depth, the converging time will not be
affected by the size of the tree. It is reasonable because every
branch of the tree is independent, and the longest branch will reach
the synchronized state at last no matter how many the branches are.
It can be seen from the compare between the trees from 10 nodes to
100 nodes from Fig.2. In the simulation, we keep the two kind of
trees with the same depth such as 1, 5 and 9.

\begin{figure}
  \center
  \includegraphics[width=8cm]{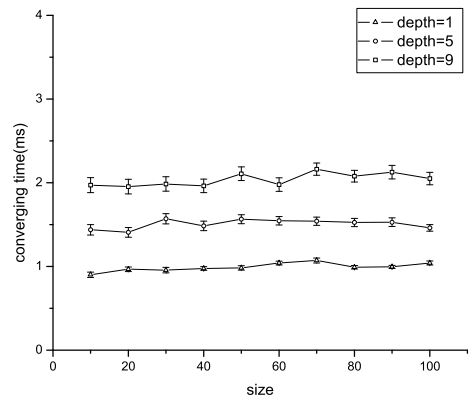}\\
  \caption{The converging time to synchronization under trees with different size. The overall coupling strength $\sigma$ is kept the same as 10. The $\triangle$ stands for the depth of the tree equals 1, the
   $\bigcirc$ stands for $depth=5$ and the $\Box$ stands for $depth=9$ .The result is averaged by
50 times.}
\end{figure}

So it is the depth of the tree that actually affects the converging
time of the synchronization process. Additional, when given the
depth of a tree, with specific oscillator model and overall coupling
strength $\sigma$, the converging time is almost a constant. Since
the initial state of each oscillator is given randomly, there may be
some fluctuation in the constant, which can be described by the
error. Thus, if we want to find a best way to distribute the weight
to enhance the synchronization, we just have to look for the
spanning tree with minimal depth. This kind of tree has minimal $R$
value and shortest converging time, which means that it has maximal
synchronizability and efficiency.

\subsection{Center of the network and the shortest spanning tree}

To get such a tree is to create a shortest oriented tree from a
undirected network. Each tree has a root, the root adopting is very
important because it determines the depth of the tree. However, the
root is not simply the node with the largest degree. For in some
case, the spanning tree created by this root may be not the
shortest. A typical example is shown in Fig.3. Here, we call the
node to create the shortest spanning tree as the center of the
network.

\begin{figure}
  \center
  \includegraphics[width=8cm]{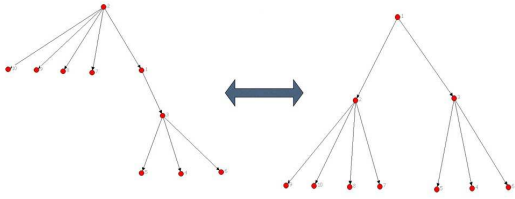}\\
  \caption{Given a network with 10 nodes, the left tree is created by adopting the node with largest degree as the root,
and the right one is the shortest tree based on this topology.
Obviously, the center is not simply the node with the largest
degree.}
\end{figure}

We use the signaling process to find the center. For a network with
$N$ nodes, every node is assumed to be a system which can send,
receive, and record signals. A node can only affect its neighbors,
which will affect their neighbors in the same way. Finally, each
node will affect the whole network. At the beginning, we set every
node as the source and give each of them one unit of signal, the
signaling process is independent. For each step, the node will
transmit the signal to the neighbors. For the signaling process is
independent, several signal can be transmitted in a node
simultaneously. After $m$ time steps, there must be a signal which
covers all the nodes in the network with the fewest steps. So
corresponding source node is the center of the network which is the
root of the shortest spanning tree.

Actually, the above signaling process could be described by a simple
but clear mathematical mechanism[25]. Suppose we have a network with
$N$ nodes, it can be represented mathematically by an adjacency
matrix $A$ with elements $A_{ij}$ equals to 1 if there is an edge
from $i$ to $j$ and 0 otherwise. And the $I$ is a matrix where all
the diagonal elements are 1 and others are 0. Then the column $i$ of
matrix $V=(I+A)^{m}$ will represent the effect of source node $i$ to
the whole network in $m$ steps. So we can get a $N$-dimensional
vector that records each node's signal quantity which represents the
effect of the source node. If all the elements of a column are
nonzero, the signal of the source node $i$ has affected the whole
network. For each step, all the columns will be updated. To find the
center, we should simply find which column reaches totally nonzero
at first.

The shortest spanning tree can be created by the signaling process
too. Suppose node $i$ is the center of the network, we mark $i$ as
the used node. The second level includes all the unused nodes
connected to the node $i$, then mark all these nodes and the edges
from $i$ to them as used. To create the third level, we consider
each node in the second level as sub-center, and the process is the
same as the center $i$. Specifically, the edges between the nodes in
the same level should be left unmarked. Moreover, if a node $k$ is
marked by one of the node in the higher level, although the node $k$
is connected to another node $j$ in the higher level, the edge
between $k$ and $j$ should be left unmarked too. After all the nodes
are marked, we can get the shortest spanning tree. The nodes in the
tree are connected by the marked edge and the direction of the edge
is from the higher level to the lower level. This spanning tree (i)
embeds a directed spanning tree, (ii) has no directed loop, and
(iii) has normalized input strengths as the Ref[20]. So the index
$R$ of the tree equals to 1. Additionally, For its depth is minimal
in all spanning trees from the given network, so the synchronization
converging time will be the shortest.

The topology of the original network is related to the depth of such
effective directed trees. For instance, the center of the scale-free
network usually has big degree. This kind of center can reduce the
depth of the tree significantly. On the contrary, the homogenous
random network does not have such kind of center, so the depth of
the spanning tree from these networks will always be longer than
that from the scale-free network as shown in Fig.4.

\begin{figure}
  \center
  \includegraphics[width=8cm]{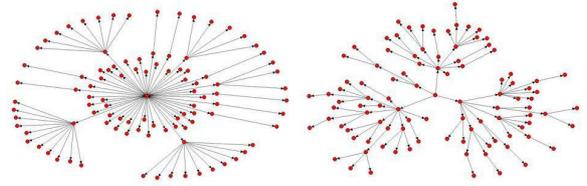}\\
  \caption{The left tree is a shortest spanning tree created from a scale-free network, which is from BA model with average degree 6.
   The right one is from a random network with the same average degree as the scale-free network. It clear that the depth of
such effective directed trees is related to the topology of the
original networks.}
\end{figure}

\section{Conclusion}

In the former works, to enhance the synchronization is always to
reduce the index $R$. That is to reduce the overall coupling
strength. However, the converging time is also an important factor.
In this paper, we find that the depth of the tree is the only factor
that effects the converging time when given the oscillator model and
the overall coupling strength $\sigma$. Additionally, we present a
simple way to obtain the shortest spanning tree with maximal
synchronizability and efficiency from any given network.

In order to enhance the synchronizability of a network, the coupling
strength of each edge is always be scaled by some structural
characteristics such as degree, betweeness and so forth. The purpose
of this process is to reduce the effect of the edges which is
disadvantage for synchronization. To make the process to an extreme
way, creating a spanning tree from a network is to cut off some
edges in the network, which can also be considered as imposing the
weight of these edges to 0. All the spanning trees from the network
have the same $R=1$, which means the critical overall couple
strength of the tree is minimal. Hence, the tree creating is a
coupling strength scaling process.

Moreover, compared with any spanning tree from a given network, the
shortest spanning tree can reduce the converging time significantly.
It makes the tree have maximal efficiency. In fact, comparing with
the former works[15-16], the way presented by us is universal. It is
valid not only in the heterogeneous networks, but also in the
homogenous networks.

\section*{Acknowledgement}
We thank professor Yin Fan and Dong Zhou for many useful
suggestions. This work is supported by  NSFC under Grants
No.70771011, No.70431002, No.60534080.


\begin{thebibliography}{999}
\bibitem{1}T. Nishikawa, A. E. Motter, Y.-C. Lai, and F. C. Hoppensteadt, Phys. Rev. Lett. \textbf{91}, 014101 (2003).
\bibitem{2}H. Hong, B. J. Kim, M. Y. Choi, and H. Park, Phys. Rev. E \textbf{69}, 067105 (2004).
\bibitem{3}P. N. McGraw and M. Menzinger, Phys. Rev. E \textbf{72}, 015101(R) (2005).
\bibitem{4}M. Zhao, T. Zhou, B.-H. Wang, G. Yan, H.-J. Yang, and W.-J.Bai, Physica A \textbf{371}, 773 (2006).
\bibitem{5}X. Wu, B.-H. Wang, T. Zhou, W.-X. Wang, M. Zhao, and H.-J. Yang, Chin. Phys. Lett. \textbf{23}, 1046 (2006).
\bibitem{6}D.J. Watts, S.H. Strogatz, Nature (London) \textbf{393}, 440 (1998) .
\bibitem{7}M. Barahona, L.M. Pecora, Phys. Rev. Lett. \textbf{89}, 054101 (2002).
\bibitem{8}H. Hong, B.J. Kim, M.Y. Choi, H. Park, Phys. Rev. E \textbf{69}, 067105 (2004).
\bibitem{9}T. Nishikawa, A.E. Motter, Y.-C. Lai, F.C. Hoppensteadt, Phys. Rev. Lett. \textbf{91}, 014101 (2003).
\bibitem{10}M.E.J. Newman, Phys. Rev. Lett. \textbf{89}, 208701 (2002).
\bibitem{11}Adilson E.Motter, Changsong Zhou, and Jurgen Kurths, Phys. Rev. E \textbf{71}, 016116 (2005).
\bibitem{12}M.Chavez, D.-U.Hwang, A.Amann, H.G.E.Hentschel, and S.Boccaletti, Phys. Rev. Lett. \textbf{94}, 218701 (2005).
\bibitem{13}Changsong Zhou,Adilson E.Motter,and Jurgen Kurths, Phys. Rev. Lett. \textbf{96}, 034101 (2006).
\bibitem{14}M.Zhao, T.Zhou, B.-H.Wang, Q.Ou, J.Ren, Eur. Phys. J. B \textbf{53}, 375-379 (2006).
\bibitem{15}D.-U. Hwang, M. Chavez, A. Amann, S. Boccaletti, Phys. Rev. Lett. \textbf{94}. 138701 (2005).
\bibitem{16}Y.-F. Lu, M. Zhao, T. Zhou, B.-H. Wang, Phys. Rev. E \textbf{76}. 057103 (2007).
\bibitem{17}M. Barahona, L.M. Pecora, Phys. Rev. Lett. \textbf{89}. 054101 (2002).
\bibitem{18}L.M. Pecora, T.L. Carroll, Phys. Rev. Lett. \textbf{80}. 2109-2112 (1998).
\bibitem{19}K.S. Fink, G. Johnson, T. Carroll, D. Mar, L. Pecora, Phys. Rev. E \textbf{61}.5080-5090 (2000).
\bibitem{20}T. Nishikawa, A.E. Motter, Phys. Rev. E \textbf{73}. 065106 (2006).
\bibitem{21}Alex Arenas, Albert Diaz-Guilera, Jurgen Kurths, Yamir Moreno and Changsong Zhou, Physics Reports \textbf{469}. 93-153 (2008).
\bibitem{22}A.T. Winfree. J. Theoret, Biol. \textbf{16}. 15-42 (1967).
\bibitem{23}S.H. Strogatz, Physica D \textbf{143}. 1-20 (2000).
\bibitem{24}J.A. Acebr¨®n, L.L. Bonilla, C.J. P¨¦rez-Vicente, F. Ritort, R. Spigler, Rev. Mod. Phys. \textbf{77}. 137-185 (2005).
\bibitem{25}Yanqing Hu, Menghui Li, Peng Zhang, Ying Fan, and Zengru Di, Phys. Rev. E \textbf{78}, 016115 (2008).
\end{thebibliography}
\end{document}